\documentclass[smallextended,refree]{svjour3}
\usepackage{bm}
\usepackage{amsmath}
\usepackage{amsfonts}
\usepackage{amssymb}
\usepackage{enumerate}

\journalname{Int J Theor Phys}

\begin{document}

\title{$s$-Parameterized Fock Space Projectors Gained via General Ordering Theorem}
\author{F. Sh\"ahandeh$^*$\thanks{$^*$Corresponding author} \and M. R. Bazrafkan \and M. Ashrafi}
\institute{F. Sh\"ahandeh \at Physics Department, Faculty of Science, I. K. I. University, Qazvin, Iran.~\email{shahandeh@ikiu.ac.ir} \and M. R. Bazrafkan \at Physics Department, Faculty of Science, I. K. I. University, Qazvin, Iran.~\email{bazrafkan@ikiu.ac.ir} \and M. Ashrafi \at Physics Department, Faculty of Science, I. K. I. University, Qazvin, Iran.~\email{ashrafi@ikiu.ac.ir}}

\maketitle

\begin{abstract}
Employing the general ordering theorem, operational methods and the incomplete 2-dimensional Hermite polynomials we have derived the $t$-ordered expansion of the Fock space projectors. Using the result, a new integration formula regarding incomplete 2-dimensional Hermite polynomials is obtained.
\end{abstract}


\section{Introduction}

Undoubtedly, one of the most important representations in quantum mechanics, is that of the number or Fock states.~\cite{Schleich} In this regard, we are interested in the quantum mechanical phase-space representation through these states. The latter, however, is not unique due to the existence of the $s$-parameterized class of orderings and quasiprobabilities.~\cite{Glauber} A recent approach to $s$-ordering of operators is given by Sh\"ahandeh and Bazrafkan~\cite{SB} by which one may $t$-order any multiplicative sequence of $s_j$-ordered functions with $j \in \left\{ {2,3, \ldots ,k} \right\}$ as
	\begin{eqnarray} \label{sjt}
	{\left\{ {\hat F\left( {a^\dag,a} \right)} \right\}_{{s_2}}}{\left\{ {\hat G\left( {a^\dag,a} \right)}\right\}_{{s_3}}}\cdots 	{\left\{ {\hat H\left( {a^\dag,a} \right)} \right\}_{{s_k}}} = \nonumber \\ 								\sum\limits_{{\text{$i$-pair $\left({{u_l},t} \right)$-contractions}}}{{\left\{ {{\left( {\hat F\hat G \cdots \hat H} \right)}_i^{\mathbf{u}}\left( {a^\dag, a} \right)} \right\}}_t}~,
	\end{eqnarray}
in which $l \in \left\{ {1,2, \ldots ,i} \right\}$, ${\mathbf{u}} \equiv \left( {{{u}_1},{{u}_2} \ldots ,{{u}_i}} \right)$ with $u_l \in \left\{ {{-1,1,s_j}} \right\}$, and ${\left( {\hat F\hat G \cdots \hat H} \right)_i^{{\mathbf{u}}}}$ is the $i$-pair $\left (u_l,t\right )$-contracted form of the multiplicative sequence $\hat F\hat G \cdots \hat H$. In this notation, the $s$-ordering symbol has been denoted by $\left\{ \cdots \right\}_s$. 

The relation \eqref{sjt} provides the most general method of ordering of operators. As a simple example, one can evaluate the $s$-ordered form of the monomial ${\left( {{a^\dag }a} \right)^2}$ as
	\begin{eqnarray*}
	{\left( {{a^\dag }a} \right)^2} &=& {\left\{ {{{\left( {{a^\dag }a} \right)}^2}} \right\}_s} + \left[ {\left( {\frac{{s + 1}}{2}} \right) +3\left( {\frac{{s - 1}}{2}} \right)} \right]{\left\{ {{a^\dag }a} \right\}_s} \nonumber \\
	&+& \left( {\frac{{s + 1}}{2}} \right)\left({\frac{{s - 1}}{2}} \right) + {\left( {\frac{{s - 1}}{2}} \right)^2}~,
	\end{eqnarray*}
which could be simply verified using the relation~\cite{Wunsche}
	\begin{eqnarray*} \label{st}
	{\left\{ {{a^{\dag n}}{{a}^m}} \right\}_{s}} = &\sum\limits_{i = 0}^{\min \left\{ {n,m} \right\}} &{\left( {\begin{array}{*{20}{c}}
	n \\ 
	i 
	\end{array}} \right)} \left( {\begin{array}{*{20}{c}}
	m \\ 
	i 
	\end{array}} \right)i!{\left[ {\left( {\frac{{t - s}}{2}} \right)} \right]^i}\nonumber \\
	&&\times{\left\{ {{a^{\dag n-i}}{{a}^{m-i}}}\right\}_t}~.
	\end{eqnarray*}

In the present paper, we use this new technique to $t$-order Fock space projectors.

\section{Incomplete 2-Dimensional Hermite Polynomials}

As will be seen later, the class of incomplete 2-D Hermite polynomials is closely related to the general ordering problem. These polynomials are defined through~\cite{Dattoli}
	\begin{equation}
	{h_{m,n}}\left( {x,y|\tau } \right) = \sum\limits_{i = 0}^{ min\left\{m,n\right\} }  {\left( {\begin{array}{*{20}{c}}
  m \\ 
  i 
\end{array}} \right)\left( {\begin{array}{*{20}{c}}
  n \\ 
  i 
\end{array}} \right)i!{\tau ^i}{x^{m - i}}{y^{n - i}}}~,
	\end{equation}
with the generating function
	\begin{equation}
	\sum\limits_{m,n = 0}^\infty  {\frac{{{\lambda ^m}{\mu ^n}}}{{m!n!}}{h_{m,n}}\left( {x,y|\tau } \right)}  = {e^{\lambda x + \mu y + \tau \lambda \mu }}~.
	\end{equation}
One may also simply check that the partial sum formulas might be written as
	\begin{eqnarray}
	\sum\limits_{m = 0}^\infty  {\frac{{{\lambda ^m}}}{{m!}}{h_{m,n}}\left( {x,y|\tau } \right)}  = {\left( {y + \tau \lambda } \right)^n}{e^{\lambda x}}~, \label{IHpar1}\\
	\sum\limits_{n = 0}^\infty  {\frac{{{\mu ^n}}}{{n!}}{h_{m,n}}\left( {x,y|\tau } \right)}  = {\left( {x + \tau \mu } \right)^m}{e^{\mu y}}~. \label{IHpar2}
	\end{eqnarray}
These functions, using the general ordering theorem (GOT) Eq.~\eqref{sjt}, readily provide the relations
	\begin{equation}
	{a^{\dag n}}{a^m} = {\left\{ {{h_{n,m}}\left( {{a^\dag },a|{\tau _ - }} \right)} \right\}_s}~,\>\ \tau _ - \equiv {\frac{{s - 1}}{2}}~, \label{aDnam}
	\end{equation}
and
	\begin{equation}
	\left\{ {{a^{\dag n}}{a^m}} \right\}_s = {\left\{ {{h_{n,m}}\left( {{a^\dag },a|\tau_{st}} \right)} \right\}_t},\>\ \tau _{st} \equiv \frac{{t - s}}{2}~. \label{aDnam_s}
	\end{equation}

\section{$t$-ordered form of $\left\{ {{e^{\lambda {a^\dag }a}}} \right\}_s$}

For future applications, we use GOT to $t$-order the operator $\left\{ {{e^{\lambda {a^\dag }a}}} \right\}_s$. To this end, we may use Eq.~\eqref{aDnam_s} to write
	\begin{eqnarray*}
	{\left\{ {{e^{\lambda {a^\dag }a}}} \right\}_s} = {\left\{ {\sum\limits_{n = 0}^\infty  {\frac{{{\lambda ^n}}}{{n!}}{h_{n,n}}\left( {{a^\dag },a|{\tau _{st}}} \right)} } \right\}_t}~,	
	\end{eqnarray*}
which having the relation to the usual Laguerre polynomials,~\cite{Dattoli,Wunsche1}
	\begin{equation} \label{rtouL}
	{h_{n,n}}\left( {x,y|\tau } \right) = {\tau ^n}n!{L_n}\left( { - \frac{{xy}}{\tau }} \right)~,
	\end{equation}
gives
	\begin{equation} \label{eLag}
	{\left\{ {{e^{\lambda {a^\dag }a}}} \right\}_s} = {\left\{ {\sum\limits_{n = 0}^\infty  {{{\left( {\lambda {\tau _{st}}} \right)}^n}{L_n}\left( { - \frac{{{a^\dag }a}}{{{\tau _{st}}}}} \right)} } \right\}_t}~.
	\end{equation}

Now, one may employ the generating function of the Laguerre polynomials~\cite{Bayin} to reduce Eq.~\eqref{eLag} to
	\begin{eqnarray}
	{\left\{ {{e^{\lambda {a^\dag }a}}} \right\}_s} = f{\left\{ e^{ga^\dag a } \right\}_t} \label{eL} \\
f \equiv \frac{2}{{2 - \lambda \left( {t - s} \right)}}, \>\ g \equiv \lambda f~.
	\end{eqnarray}

\section{$t$-ordering of Fock Space Projectors}

In this section we give a convenient way of $t$-ordering the Fock states projectors using GOT in combination with the operational methods in generatingfunctionology.~\cite{Wilf} We may begin with the well-known relation $\left| 0 \right\rangle \left\langle 0 \right| = :{e^{ - {a^\dag }a}}:$.~\cite{Louisell} Now, we may write the projectors in the form
	\begin{eqnarray}
  \left| n \right\rangle \left\langle m \right| &=& \frac{1}{{\sqrt {n!m!} }}{a^{\dag n}}\left| 0 \right\rangle \left\langle 0 \right|{a^m} \nonumber \\ 
&=& \frac{1}{{\sqrt {n!m!} }}{a^{\dag n}}:{e^{ - {a^\dag }a}}:{a^m}
	\end{eqnarray}

At this point, we may use GOT to $t$-order the product ${a^{\dag n}}{\left\{ {{e^{\lambda {a^\dag }a}}} \right\}_s}{a^m}$. First, we have
	\begin{eqnarray*}
	{a^{\dag n}}{\left\{ {{e^{\lambda {a^\dag }a}}} \right\}_s} &=& {\left\{ {\sum\limits_{m = 0}^\infty  {\frac{{{\lambda ^m}}}{{m!}}{a^{\dag m}}\sum\limits_{i = 0}^{\min \left\{ {n,m} \right\}} {\left( {\begin{array}{*{20}{c}}
  n \\ 
  i 
\end{array}} \right)\left( {\begin{array}{*{20}{c}}
  m \\ 
  i 
\end{array}} \right)i!{{\left( {\frac{{s - 1}}{2}} \right)}^i}{a^{\dag n - i}}{a^{m - i}}} } } \right\}_s} \nonumber \\ 
&=& {\left\{ {\sum\limits_{m = 0}^\infty  {\frac{{{\lambda ^m}}}{{m!}}{a^{\dag m}}{h_{n,m}}\left( {{a^\dag },a|{\tau _ - }} \right)} } \right\}_s}~,
	\end{eqnarray*}
which after using the partial sum formula for incomplete 2-D Hermite polynomials Eq.~\eqref{IHpar2} gives
	\begin{equation} \label{aDe}
	{a^{\dag n}}{\left\{ {{e^{\lambda {a^\dag }a}}} \right\}_s} =  {\left( {\lambda {\tau _ - } + 1} \right)^n}{\left\{ {{a^{\dag n}}{e^{\lambda {a^\dag }a}}} \right\}_s}~.
	\end{equation}

Again, we may multiply the right-hand-side of Eq.~\eqref{aDe}, regardless of the constant coefficient, on the right by $a^m$. This leads to
	\begin{eqnarray*}
	{\left\{ {{a^{\dag n}}{e^{\lambda {a^\dag }a}}} \right\}_s}{a^m} &=& \left\{ {\sum\limits_{k = 0}^\infty  {\frac{{{\lambda ^k}}}{{k!}}{a^k}\sum\limits_{j = 0}^{\min \left\{ {k + n,m} \right\}} {\left( {\begin{array}{*{20}{c}}
  {k + n} \\ 
  j 
\end{array}} \right)\left( {\begin{array}{*{20}{c}}
  m \\ 
  j 
\end{array}} \right)j!} } } \right. \nonumber \\ 
&& \qquad {\left. {{{\left( {\frac{{s - 1}}{2}} \right)}^j}{a^{\dag k + n - j}}{a^{m - j}}} \right\}_s} \nonumber \\
&=& {\left\{ {{{\left( {{\tau_{-}\partial _{{a^\dag }}} + a} \right)}^m}{a^{\dag n}}{e^{\lambda {a^\dag }a}}} \right\}_s}
	\end{eqnarray*}
Using the substitution $a^{\dag n} \to a^{-n}{\partial _\lambda ^n}$ we arrive at
	\begin{eqnarray}
	{\left\{ {{a^{\dag n}}{e^{\lambda {a^\dag }a}}} \right\}_s}{a^m} = {\left\{ {{a^{m - n}}\partial _\lambda ^n{\kappa^m}{e^{\lambda {a^\dag }a}}} \right\}_s} \\
\kappa \equiv \tau_{-}\lambda  + 1
	\end{eqnarray}
which after changing the order of differentiation with $\kappa$ by the general Leibniz rule
	\begin{equation*}
	\frac{{{d^n}}}{{d{x^n}}}\left[ {f\left( x \right)g\left( x \right)} \right] = \sum\limits_{i = 0}^n {\left( {\begin{array}{*{20}{c}}
  n \\ 
  i 
\end{array}} \right)\frac{{{d^i}f\left( x \right)}}{{d{x^i}}}\frac{{{d^{n - i}}g\left( x \right)}}{{d{x^{n - i}}}}}
	\end{equation*}
leads to
	\begin{equation} \label{ea}
	 {\left\{ {{a^{\dag n}}{e^{\lambda {a^\dag }a}}} \right\}_s}{a^m} = \kappa^m {\left\{ {{h_{n,m}}\left( {{a^\dag }, a|{\tau _ - \kappa^{-1} }} \right){e^{\lambda {a^\dag }a}}} \right\}_s}
	\end{equation}
Combining Eq.~\eqref{aDe} with~\eqref{ea} gives
	\begin{equation} \label{aDeafinal}
	{a^{\dag n}}{\left\{ {{e^{\lambda {a^\dag }a}}} \right\}_s{a^m}} = {\kappa^{n+m}}{\left\{ {{h_{n,m}}\left( {{a^\dag }, a|{\tau _ - \kappa^{-1}}} \right){e^{\lambda {a^\dag }a}}} \right\}_s}
	\end{equation}
Equation~\eqref{aDeafinal} might be used together with Eq.~\eqref{eL} leading to
	\begin{eqnarray}
	{a^{\dag n}}{\left\{ {{e^{\lambda {a^\dag }a}}} \right\}_s{a^m}} = f{\kappa '^{n+m}}{\left\{ {{h_{n,m}}\left( {{a^\dag }, a|{\tau '_ - \kappa '^{-1}}} \right){e^{g {a^\dag }a}}} \right\}_t} \\
\tau '_{-} \equiv \frac{t - 1}{2}, \qquad \kappa ' \equiv \tau'_{-}g  + 1 
	\end{eqnarray}
This is the most general transformation of this kind. Thus, one may choose $s=1$ and $\lambda=-1$ in Eq.~\eqref{aDeafinal} to get to
	\begin{eqnarray} \label{nmproj}
	\left| n \right\rangle \left\langle m \right| = \frac{1}{{\sqrt {n!m!} }}{f^{n + m + 1}}{\left\{ {{h_{n,m}}\left( {{a^\dag },a\left| \kappa  \right.} \right){e^{ - f{a^\dag }a}}} \right\}_t} \\
f = \frac{2}{{t + 1}},\qquad \kappa  = \frac{{{t^2} - 1}}{4}
	\end{eqnarray}
which is the desired result. In the special case of $m=n$ this gives~\cite{Fan}
	\begin{equation}
	\left| n \right\rangle \left\langle n \right| = {f^{2n + 1}}{\kappa ^n}{\left\{ {{L_n}\left( { - \frac{{{a^\dag }a}}{\kappa }} \right){e^{ - f{a^\dag }a}}} \right\}_t}
	\end{equation}
where we have used the relation to the usual Laguerre polynomials Eq.~\eqref{rtouL}.

\section{An Application}

The most obvious application of Eq.~\eqref{nmproj} is to write the $\left(-t\right)$-parameterized quasiprobability of the Fock space projectors as~\cite{Glauber}
	\begin{equation}
	{W_{\left| n \right\rangle \left\langle m \right|}}\left( {\alpha , - t} \right) = \frac{1}{{\sqrt {n!m!} }}{f^{n + m + 1}}{h_{n,m}}\left( {{\alpha ^ * },\alpha \left| \kappa  \right.} \right){e^{ - f{{\left| \alpha  \right|}^2}}}
	\end{equation}
and thus, the matrix elements of any given operator $\hat F$ in Fock space representation is given by
	\begin{equation}
	\left\langle m \right|\hat F\left| n \right\rangle  = {\text{Tr}}\left\{ {\hat F\left| n \right\rangle \left\langle m \right|} \right\} = \int {\frac{{{d^2}\alpha }}{\pi }{W_{\left| n \right\rangle \left\langle m \right|}}\left( {\alpha , - t} \right){W_{\hat F}}\left( {\alpha ,t} \right)}
	\end{equation}
in which ${{W_{\hat F}}\left( {\alpha ,t} \right)}$ is the $t$-parameterized symbol function of the operator $\hat F$. Choosing $\hat F = \left| \beta  \right\rangle \left\langle \beta  \right|$ and using the $\left(-t\right)$-ordered expansion of the coherent state projectors~\cite{Fan1}
	\begin{equation}
	\left| \beta  \right\rangle \left\langle \beta  \right| = \frac{2}{{1 - t}}{\left\{ {\exp \left[ {\frac{{ - 2}}{{1 - t}}\left( {{\beta ^ * } - {a^\dag }} \right)\left( {\beta  - a} \right)} \right]} \right\}_{ - t}}
	\end{equation}
leads to
	\begin{equation} \label{b*nbm}
	{\beta ^{ * n}}{\beta ^m} = \frac{{2f^{n+m+1}{e^{{{\left| \beta  \right|}^2}}}}}{{1 - t}}\int {\frac{{{d^2}\alpha }}{\pi }{h_{n,m}}\left( {{\alpha ^ * },\alpha |{\kappa}} \right){e^{-f{{\left| \alpha  \right|}^2} - \frac{2}{{1 - t}}{{\left| {\beta  - \alpha } \right|}^2}}}}
	\end{equation}
which is a new integration formula for incomplete 2-D Hermite polynomials.

\section{Conclusion}

The GOT is a purely combinatorial  approach to the ordering problem of operators. In this regard, we have used it to give the general ordered form of the Fock space projectors. We have used the simple operational methods together with the new incomplete 2-D Hermite polynomials to achieve this aim. A simple application of the result has been given which leaded to the integration formula~\eqref{b*nbm} for incomplete 2-D Hermite polynomials.

\end{document}